\documentclass{aa}

\usepackage{amsmath}
\usepackage{slashbox}
\usepackage[pdftex]{graphicx}

\usepackage{color}

\usepackage{natbib}

\usepackage[colorlinks=True,linkcolor=blue,urlcolor=blue,citecolor=blue]{hyperref}

\begin{document}

\title{Deflagration to detonation transition by amplification of acoustic waves in type Ia supernovae}

\titlerunning{DDT by amplification of acoustic waves in type Ia supernovae}

\author{C. Charignon
      \inst{1}
      \and
      J.-P. Chi\`eze\inst{1}
      }

\institute{CEA, IRFU, SAp, 91191 Gif-sur-Yvette, France\\
          \email{camille.charignon@cea.fr}
          }

\abstract{}{We study a new mechanism for deflagration to detonation transition in thermonuclear supernovae (SNe Ia), based on the formation of shocks by amplification of sound waves in the steep density gradients of white dwarfs envelopes. We characterize, in terms of wavelength and amplitude, the perturbations which will ignite a detonation after their amplification.}{This study was performed using the well tested HERACLES code, a conservative hydrodynamical code, validated in the present specific  application by an analytical description of the propagation of sound waves in white dwarfs. Thermonuclear combustion of the carbon oxygen fuel was treated with the $\alpha$-chain nuclear reactions network.}{In planar geometry we found the critical parameter to be the height of shock formation. When it occurs in the inner dense regions ($\rho>10^6$ $g.cm^{-3}$) detonation is inevitable but can take an arbitrarily long time. We found that ignition can be achieved for perturbation as low as Mach number: M$\sim$ 0.
005, with heating times 
compatible with typical explosion time scale (a few seconds). On the opposite no ignition occurs when shocks initiated by small amplitude or large wavelength form further away in less dense regions. We show finally that ignition is also achieved in a spherical self-gravitating spherical model of cold C+O white dwarf of 1.430 $M_\odot$, but due to the spherical damping of sound waves it necessitates stronger perturbation (M$\sim$ 0.02). Small perturbations (M$\sim$ 0.003) could still trigger detonation if a small helium layer is considered. In the context of SNe Ia, one has to consider further the initial expansion of the white dwarf, triggered by the deflagration, prior to the transition to detonation. As the star expands, gradients get flatter and ignition requires increasingly strong perturbations. }{}


\maketitle

\section{Introduction}

Type Ia supernovae are thought to result from the thermonuclear explosion of a carbon oxygen white dwarf. Presently two main families of models are proposed: in the single degenerate scenario \citep{1973ApJ...186.1007W} explosion is triggered by continuous accretion of the hydrogen or helium rich envelop of a non-degenerate companion until Chandrasekhar mass is reached; in the double degenerate scenario \citep{1984ApJS...54..335I} the explosion results of the merging of two C+O white dwarfs. In the latter mechanism, only the detonation mode of combustion can release enough energy to prevent a collapse and produce a healthy explosion. Furthermore ignition is not so easy and require violent mergers \citep{2012ApJ...747L..10P}, that might not be frequent enough to explain the SN Ia rates. In the former scenario pure detonation and pure deflagration fail to reproduce light curves and nucleosynthesis. Detonation models \citep{1969Ap&SS...5..180A} incinerate the whole star at high density producing mostly $^{56}
Ni$ and resulting in wrong nucleosynthesis. On the other hand pure deflagration is very subsonic and gives time for expansion so that burning also occurs at lower density and produce the intermediate elements needed for spectra, but those models are under-energetics \citep{2007ApJ...668.1132R}. Accordingly the delayed detonation scenario has been proposed \citep{1991A&A...245..114K} to combine the advantages of both combustion modes. In this model a flame born as a subsonic deflagration is accelerated by turbulence and transitions at some point to a detonation. The initial phase allow the star to expand so that IME are produced in the final detonation stage.

However, after many years of studies the mechanism of deflagration-to-detonation transition (DDT) is still unclear. It has been shown by 3D simulations  \citep{2005ApJ...623..337G} that this mechanism was the most promising path for single degenerate models to reasonably account for observables, however due to the lack of theoretical understanding of this transition, the turning point between deflagration and detonation still relies on  educated guesses \citep{2009ApJ...704..255W} or is adjusted  in order to best reproduces nucleosynthesis of normal type Ia supernovae (\cite{2010ApJ...720...99J}). On earth, DDT has been observed in confined media, for which \cite{2010DokPh..55..480I} provided numerical evidence that the transition may occur through shock-flame interactions, see also \cite{Khokhlov1999323} and \cite{Oran20074} for more details on shock-flame interaction in DDT. See also \cite{Ciccarelli2008499} for a review of DDT mechanisms in typical terrestrial systems. However in unconfined media, 
terrestrial experiments showed that a turbulent flame is rather stable against DDT. Thus it has been proposed that the transition should occur 
through the 
gradient mechanism \citep{1991A&A...246..383K}, where a suitable gradient of the reactions induction time is created and maintained over a time long enough for a coherent burning wave to build up pressure and create a self-sustained detonation. It has been suggested that the turning point from turbulent deflagration to detonation might be the creation of such a flat gradient at the transition between the flamelet and the distributed combustion regimes, for which the flame changes in nature. However it is not clear if the turbulence level necessary for the distributed regime to occur is achieved during the explosion, even at the late stages. More recently, through numerical studies \cite{2011PhRvL.107e4501P} claimed that intense enough turbulence could directly initiate DDT, without relying on this induction time gradient. Other mechanisms able to ignite a delayed detonation from a turbulent deflagration have also been proposed: creation of hot spot, or shock convergence in the turbulent flame brush \citep{
2012ApJ...752...89K}. Thus in the context of supernova where the plasma is fully unconfined but where high level of turbulence should be present, the question is still debated.

Here we examine a new mechanism by which pressure perturbations, created by flame turbulence, are amplified during their propagation ahead of the flame in the steep density gradient where they turn into shocks, which can be strong enough to ignite a detonation. It is known since \cite{1955IAUS....2..121L} that compressible turbulence can produce sound waves, even with small rms Mach number turbulence. Moreover the turbulent combustion is likely to enhance sound production. It is also known that sound waves propagating through steep density gradients get amplified, and given a large enough density jump they result in strong shocks and substantial heating. The mechanism of formation of strong shocks in density gradient has been proposed, for example, for the heating of the solar chromosphere (see \cite{1970SoPh...12..403U} for a review).


In this paper, we analyse the propagation properties of finite amplitude pressure waves in C+O degenerate matter, first in terms of the density gradient and then in a more realistic spherical WD model. We determine the frequencies and amplitudes which lead to a successful detonation. We postpone to a following paper the study of the actual source of pressure perturbations. Our paper is constructed as follows: in the first part we present our numerical methods and the physics included. In the second part we derive an approximate equation for shock formation and amplification in density gradients in C+O degenerate matter and compare numerical results obtained for the planar models to validate the numerical simulations. Then we apply this to spherical models closer to the reality: first a cold and dense white dwarf and then a set of pre-expanded structures. Finally we briefly discuss the applicability of this mechanism to thermonuclear supernovae.

\section{Numerical Methods and Physical Input}
\label{Inputs}

We will consider the dense plasma of WD interiors composed of carbon - oxygen mixture. At these densities the plasma is degenerate and needs special treatment. Subsection \ref{code} describes the method used to simulate gas hydrodynamic in the dense plasma with the HERACLES code\footnote{\url{http://irfu.cea.fr/Projets/Site\_heracles/index.hmtl}} \citep{2007A&A...464..429G}. Subsection \ref{eos} describes the implementation of the degenerate equation of state (EoS) within the code. Finally subsection \ref{nuclear} presents the implementation of the thermonuclear reaction network.

\subsection{The Code: HERACLES}
\label{code}

The simulations are performed using the HERACLES code. It is a grid-based conservative compressible hydro code, with a second order Godunov scheme, which solves the usual Euler fluid equations:

\begin{center}
 \begin{eqnarray}
 \frac{\partial \rho}{\partial t}         + \vec{\nabla}(\rho \vec{u}) &= 0 \nonumber \\
  \frac{\partial \rho \vec{u}}{\partial t} + \vec{\nabla}(\rho \vec{u}\otimes\vec{u} + P) &= -\rho\vec{g} \nonumber \\
  \frac{\partial E}{\partial t}            + \vec{\nabla}[(E+P)\vec{u}] &= \epsilon_{nuc},
\end{eqnarray}
\end{center}

\noindent where $\vec{g}$ is the acceleration of gravity, $\epsilon_{nuc}$ is the energy release due to thermonuclear reactions and $E=E_0 + E_{th} + 0.5\rho\vec{u}^2$ is the total energy including respectively degeneracy, thermal and kinetic energy. The fluxes between cells are computed through a Riemann solver adapted to our degenerate EoS. It also include gravity implemented as a Cranck-Nicholson step decoupled from the hydro step. Since we made only 1D calculations, the gravitational acceleration is either constant in the planar geometry or is computed self-consistently at every radius $r$ as the acceleration of gravity due to the enclosed mass:

$$ g(r)= -\frac{Gm(r)}{r^2}, $$   

\noindent where  $m(r)=\int_0^r 4\pi r^2\rho(r)dr$ is the enclosed mass in the sphere of radius $r$ and G is gravitational constant.

\subsection{Equation of State}
\label{eos}

In a white dwarf, the electron gas is degenerate and ions form an almost perfect gas. Depending on the temperature and density, electrons have various degrees of degeneracy. Thus a general EoS is used to described the plasma including a perfect gas of ions, photons in local thermal equilibrium and electrons at all degrees of degeneracy and relativity \citep{2000ApJS..126..501T}. To solve the Riemann problem for a general EoS we use the solver described by \cite{1985JCoPh..59..264C} as implemented in the FLASH code \citep{2000ApJS..131..273F}. But in some cases it was not accurate enough to follow precisely the temperature. Indeed in the density and temperature conditions relevant here, the main contribution to the pressure and energy of the gas is from the degenerate electrons and is essentially independent of the temperature. The thermal contribution, defined as $E_{th}(\rho,T)=E_{in}(\rho,T)-E_{in}(\rho,T=0 K)$ represents often less than 0.1\% of the internal energy ($E_{in}$). Since a conservative code 
computes 
only the total energy: $E=E_0 + E_{th} + 0.5\rho\vec{u}^2$, direct inversion to obtain the temperature is sometimes inaccurate. Thus, in order to accurately follow the temperature of the plasma, an equation of evolution for the thermal energy is derived and used when necessary:

$$ \frac{\partial E_{th}}{\partial t} + \nabla (\vec{u} E_{th}) = - P_{th} \nabla \vec{u}  $$

\noindent where $P_{th}$ is the thermal pressure defined as $P_{th}(\rho,T)=P(\rho,T)-P(\rho,T=0 K)$.

\subsection{Numerical Dissipation and Waves Propagation}

In order to resolve sound waves with minimum numerical dissipation, there must be enough points per wavelength and the time step must be much smaller than the period. To check the ability of our code to propagate sound waves without too much dissipation, we initiate waves of different frequencies at a given amplitude and measure the amplitude at the opposite side. With a 1500 points grid the numerical losses are only 5\% for a wavelength of 1/30 of the computational box length , which correspond to 50 points per period. For larger wavelength the losses are smaller (2\% at 1/20 and less than 1\% at more than 1/10). As will be shown latter, spatial frequencies higher than 1/20 are not relevant to the problem considered here, thus the code is actually well suited for the present study.

\subsection{Nuclear Network and Detonations}
\label{nuclear}

\begin{figure}
 \begin{center}
 \includegraphics[width=0.45\textwidth]{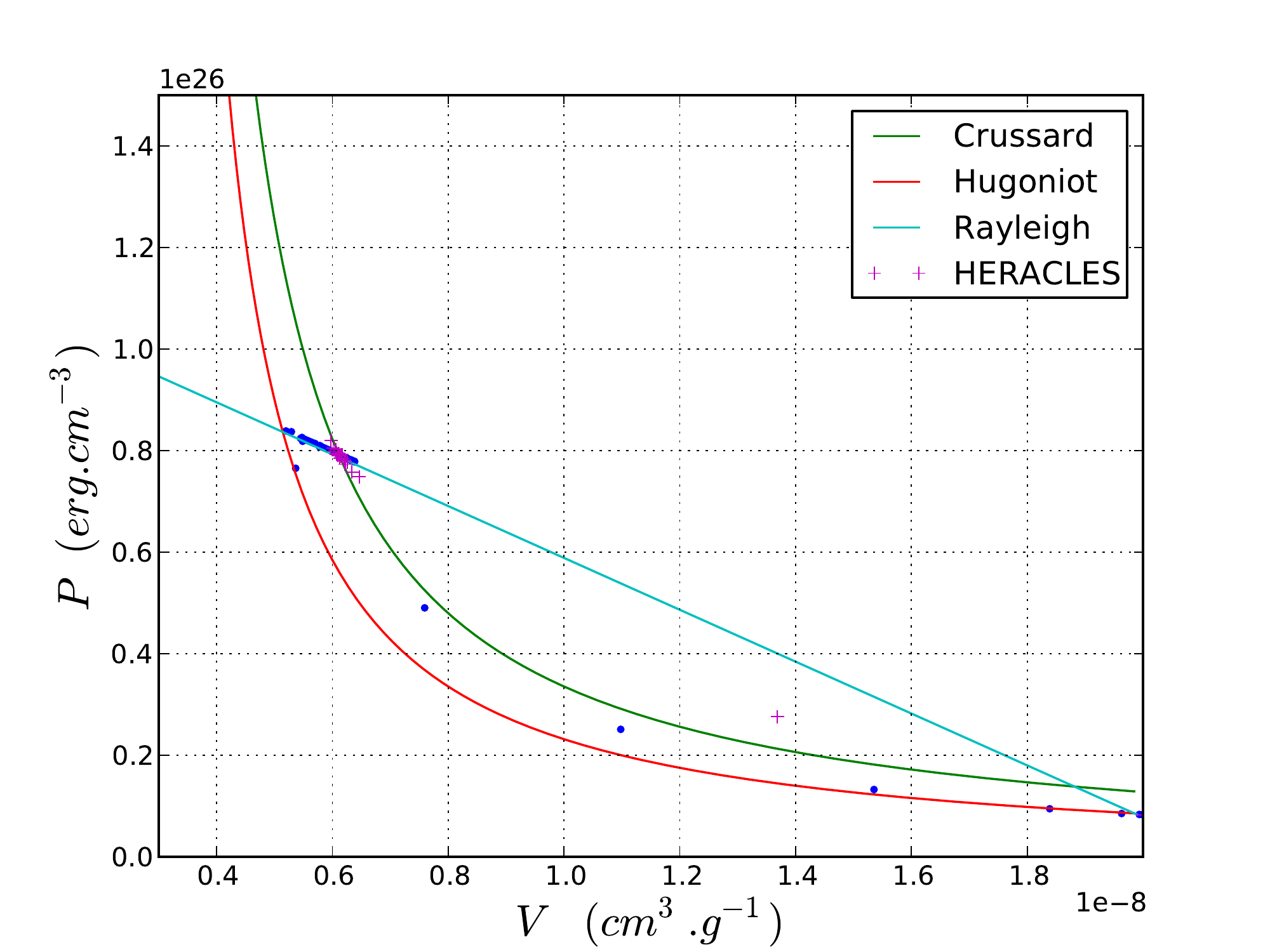}
 \caption{Pressure ($P$)-specific volume ($V$) diagram of a forced detonation in a plasma of density $\rho_0 = 5.10^7$ $g\:cm^{-3}$. The Hugoniot curve represents the energy conservation through a shock, the Rayleigh line is the line of mass and momentum conservation and the Crussard curve corresponds to the energy conservation after the nuclear burning is complete. Each blue point is the ($P$,$V$) value of a simulation cell from the high resolution code, ASTROLABE. This simulation resolves the reaction zone following the leading shock, while the crosses representing the simulation cells of the code HERACLES, jump directly to the final state. This demonstrates that even without resolving the combustion zone, this code reproduces the correct energetics and dynamics. }
 \label{Crussard}
\end{center}
\end{figure}

Since we are not concerned here with the details of nuclear burning, we adopt the classical $\alpha$ chain network, composed of 13 nuclei from $^{12}C$ to $^{56}Ni$ plus $^{4}He$. This time saving network is sufficient to describe the energetics of the C+O combustion \citep{1999ApJS..124..241T}. For the reaction rates we use the NACRE compilation. The implementation of nuclear reactions in the hydro code HERACLES, uses a decoupled combustion step with multiple local nuclear time steps. To check this implementation we compare forced detonations with a high resolution adaptive mesh code, which can nicely resolve flames and detonations (ASTROLABE code, developed by one of us: J.-P. Chi\`eze). This is visible on Fig.(\ref{Crussard}), where the cells from ASTROLABE are plotted in blue dots along the Rayleigh line for mass and momentum conservation, the Rankine-Hugoniot curve for energy conservation and the Crussard curve (the Crussard curve is the non-adiabatic Rankine-Hugoniot curve, modified by 
the energy released by the combustion). We can see, according to the ZND model, that the detonation results of a primary shock which heats up and 
compresses the plasma, pushing it on the Hugoniot. Then, combustion on scales much larger than the shock thickness, will expand the plasma and reach the intersection of the Crussard curve and the Rayleigh line. For the same set up, we plotted the simulation cells for the HERACLES code. Due to lower resolution, the initial state at $v=2\:10^{-8}\:cm^3\:g^{-1}$ goes almost directly to the final state where the combustion is complete. But even without resolving the combustion zone, the final state corresponds nicely with the high resolution simulation . Then we compare several detonations from HERACLES and found detonation speeds agreeing within less than 1\%. Finally we consider the effect of resolution on detonation initiation. The mechanism described later is self-consistent and rely only on compression and heating by shocks. There is no artificial criteria or threshold for ignition. Usually only one cell runs away, when the density and temperature of the cell are sufficient so that heating 
due to combustion is faster than the cooling following the shock. In those conditions the cell reaches sufficiently high pressure to compress the neighbouring cells, propagating the detonation. This means that the ignition scale is not resolved. But since density and temperature are averaged over the cell, the under-resolved ignition conditions are a lower estimate. Indeed, ignition occurs in the region of decompression following the shock, where temperature and density decrease almost linearly from the post shock state. Thus inside the detonating cell, the unresolved region of maximum density and temperature would runaway faster than the whole cell, with averaged density and temperature.

\section{Shock formation and heating mechanisms}

In this section we analyse the propagation and the amplification of sound waves in slab geometry. We consider a vertical column, with $h$ measuring  the height from the base ($h=0$) to the top ($h=L$). The gravity points toward the base. By an analogy to spherical geometry, we speak there of inner boundary and upper density, while  the top of the column would correspond to the outer boundary and lower densities. We adopt this simplifying assumption in order to derive a simple analytical description which is then compared to numerical simulations of hydrodynamics to validate them.

\subsection{Analytical description}

The Poynting flux of a wave of velocity amplitude  $u$ is $F=\frac{1}{2}\rho u^2 C_s$. This corresponds to the energy flux carried by the wave in the linear regime. This flux is conserved in the absence of dissipation. If the perturbation originates from a region with density $\rho_0$ and sound speed $ C_{s,0} $, its amplitude at any height $h$ reads as:

\begin{equation}
  u(h) = u_0 \sqrt{\frac{\rho_0 C_{s,0}}{\rho(h) C_s(h)}}.
 \label{Ampli_U}
\end{equation}

\noindent In degenerate matter the sound speed depends mainly on the density. Thus a wave travelling outward will slow down, so that the perturbation amplitude will grow to conserve the flux. And this, until a shock forms and dissipates kinetic energy. A good way of estimating the height of shock formation, $H_{sh}$, is to follow the hills of the wave until they catch up the valleys. The hills move with respect to the wave at $+u(h)$ while the valleys at $-u(h)$ and they will merge when they have travelled their initial separation of $\lambda_0/2$. For a perturbation initiated at $h=0$, this reads as:

\begin{equation}
 \int_0^{H_{sh}} u_0 \sqrt{\frac{\rho_0 C_{s,0}}{\rho C_s}}\frac{dh}{C_s} = \frac{\lambda_0}{4},
 \label{Hsh}
\end{equation}

\begin{figure}
 \begin{center}
 \includegraphics[width=0.45\textwidth]{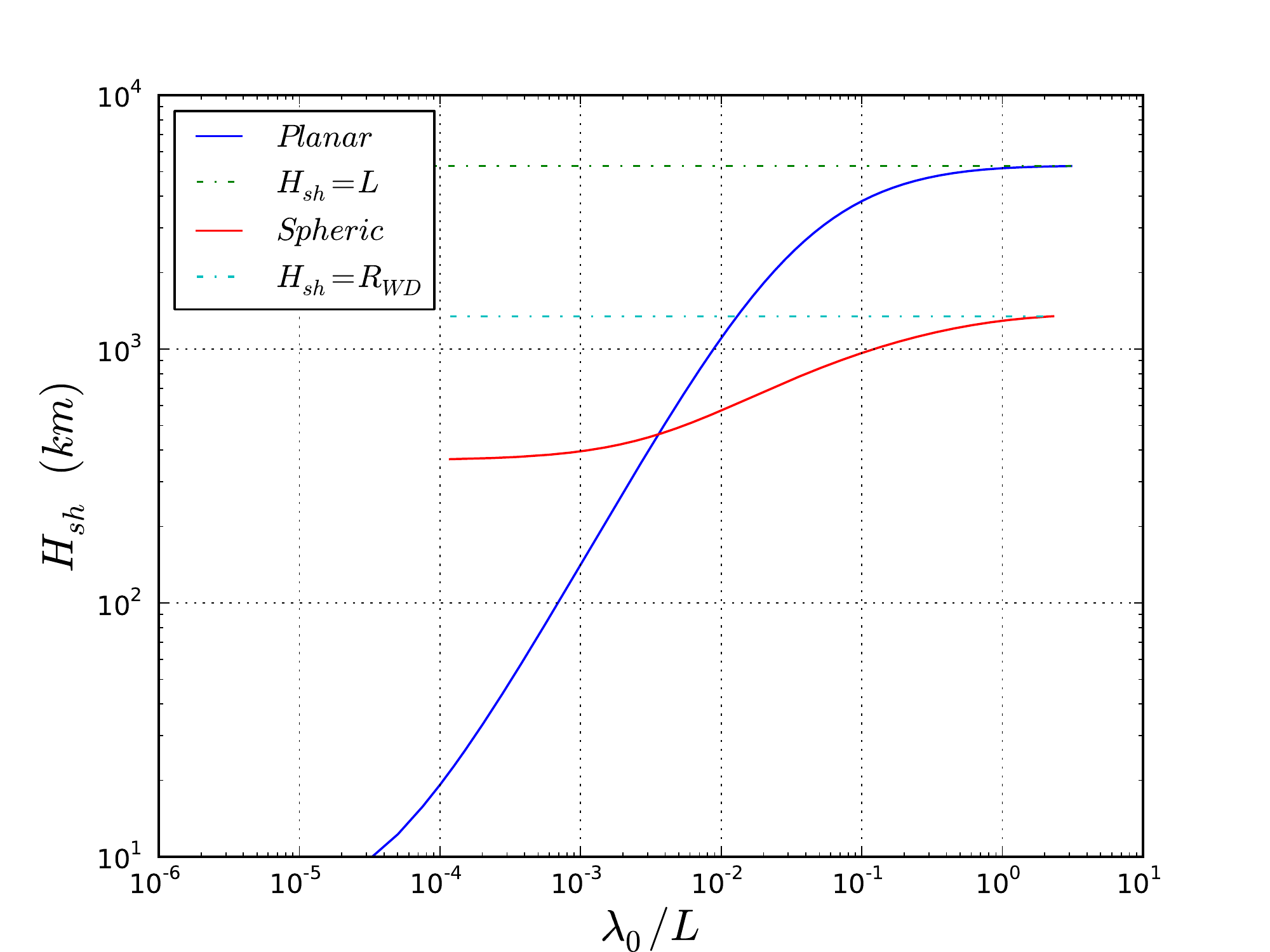}
 \caption{Height of shock formation versus the initial wavelength of the perturbation normalised to the size of the box: $\lambda_0/L$ in the planar and spherical cases.}
 \label{ShockFormation}
\end{center}
\end{figure}

\noindent The height of shock formation is calculated for a barometric structure with an upper density of $\rho_0=5.10^9$ $g\:cm^{-3}$ decreasing to a density of $10^4$ $g\:cm^{-3}$, over a length of $L=5260$ $km$. The choice of the lower density is a limit where thermonuclear ignition would no longer be possible. The constant gravity acceleration is: $ g = GM_{\odot}/(1500km)^2 $ and the sound speed at $h=0$ is $C_{s,0}=10500$ $km\:s^{-1}$. Equation (\ref{Hsh}) for this gradient is plotted versus the frequency in Fig.\ref{ShockFormation}. It provides a quite good estimate of the height at which a perturbation starts to dissipate and heat up the medium, due to the formation of shocks. There are two trends: the longer the wavelengths the farther they start dissipating and for a given wavelength $\lambda_0 $, stronger initial perturbation starts dissipating earlier.
$H_{sh}$ determines if heating will occur in regions where densities are sufficient for ignition to occur, when $H_{sh}<L$. In that case the heating rate is increasing with the perturbation amplitude $u$ and the detonation initiation only necessitates a sufficient number of shocks to reach ignition temperature. On the other hand when $H_{sh}>L$, shocks form too late in regions of low densities, where detonation is not possible whatever the number of shocks. 

%

The amount of energy dissipated by a shock depends on its strength defined as $\eta=(\rho^*-\rho)/\rho$, with $\rho^*$ and $\rho$ the shocked and pre-shock densities respectively. To estimate the shock strength we follow \cite{1967ZA.....67..193U} and \cite{1970SoPh...12..403U}. Since we consider very weak perturbations the resulting shocks are never strong ($\eta \sim 1$), this implies they will be of the sawtooth type (\cite{1959flme.book.....L} \S 102 p385). For this shape the energy carried by the 
wave is:

$$ F = \frac{1}{T} \int_0^T (P-P_0)udt = \frac{1}{12} \Delta P \Delta u ,$$

\noindent where $\Delta P$ and $\Delta u$ are the jumps in pressure and velocity across the shock front and $T$ is the period. Assuming weak shocks, $\Delta u=\eta C_s$ and $\Delta P = \rho C_s^2 \eta$ and using the adiabatic index $\gamma=1+P/E$ and $C_s^2=\gamma P/\rho$, this formula reduces to:

\begin{equation}
 F = \frac{1}{12} \gamma P_0 C_s \eta^2.
\label{FluxEnergy}
\end{equation}

\noindent  On the other hand, shocks dissipate kinetic energy into internal energy. The corresponding rate can be estimated following \cite{1959flme.book.....L} for weak shocks, or for arbitrary strong shocks as in \cite{1947PhRv...72.1109B}:

\begin{equation}
 \frac{1}{F}\frac{dF}{dh} = \frac{\gamma+1}{C_s}\eta \nu,
\label{Dissipation}
\end{equation}

\noindent where $\nu=1/T$ is the frequency. Then deriving eq.(\ref{FluxEnergy}) and equating it to dissipation losses, we get the shock strength equation:

\begin{equation}
 \frac{d\eta}{dh} = \frac{\eta}{2} \left( -\frac{1}{\gamma} \frac{d\gamma}{dh} + \frac{\gamma g}{C_s^2}  - \frac{1}{2C_s^2}\frac{dC_s^2}{dh} -\frac{\gamma+1}{C_s}\eta \nu \right).
\label{EquEta}
\end{equation}

\noindent \cite{1961ApJ...134..347O} showed that for weak shock the conserved quantity is actually $F C_s^2$. The inclusion of this refraction effect adds an additional factor $\frac{1}{C_s^2}\frac{dC_s^2}{dh} $. Eq.(\ref{EquEta}) can then be integrated to get shock strengths $\eta(h)$ given the position of formation and an initial shock strength $\eta_0$. We note that the pressure term, $\gamma g/C_s^2$, and the energy dissipation rate, $ (\gamma+1)\eta \nu/C_s $, dominate the RHS of equation (\ref{EquEta}), so that an asymptotic behaviour of the shock strength can be approximated by equating these two terms to obtain:

\begin{equation}
 \eta_\infty = \frac{\gamma g}{(\gamma+1)C_s \nu}.
 \label{EtaAsymptotic}
\end{equation}

\noindent This relation is plotted as solid lines on Fig.\ref{Parametric} for different frequencies or equivalently initial wavelengths. It can be noticed that the asymptotic value depends only on the frequency of the perturbation. Accordingly perturbations of given wavelength with different amplitudes tend to a common asymptotic behaviour: $\eta_{\infty,\nu}(h)$.

%

\subsection{Numerical description and validation}

Here we use (\ref{EquEta}) to validate the numerical code used in the following section, which will be devoted to ignition. In our simulations, to create the initial sound waves we impose oscillating density and momentum perturbations at the inner eulerian boundary, $h=0$: 

\begin{equation} 
 \left\{
\begin{array}{l l}
\rho(h=0)   &= \rho_0(1-u_0 sin(2\pi\nu t)/C_{s,0}), \\
\rho u(h=0) &= -\rho_0{u_0}sin(2\pi\nu t). \\
\end{array} \right.
\label{CL_L}
\end{equation}

\begin{figure}
 \begin{center}
 \includegraphics[width=0.45\textwidth]{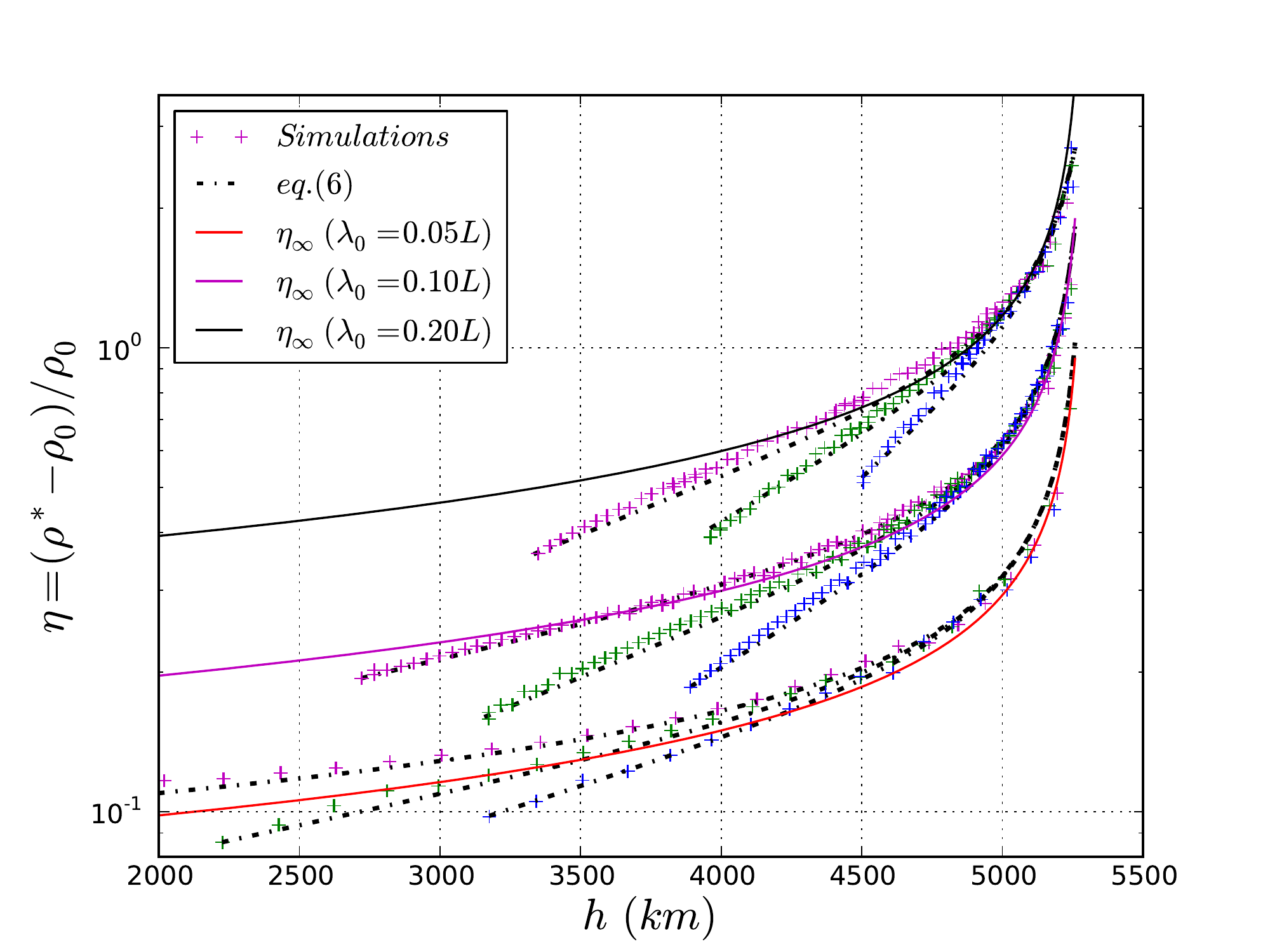}
 \caption{Profiles of shock strength. Dots represent the measurements from our numerical simulations: three different amplitudes (200, 100, 50 $km\:s^{-1}$) in magenta, green and blue respectively for each wavelength families represented by their common asymptote eq.(\ref{EtaAsymptotic}) for the three wavelengths ($\lambda_0=0.05$L, 0.1L, 0.2L in solid lines in red, magenta and black respectively). Also plotted are the predictions from eq.(\ref{EquEta}) in black dashed lines.}
 \label{Parametric}
\end{center}
\end{figure}

\noindent The velocity perturbation ($u_0$) and the wavelength ($\lambda_0 =  C_{s,0} /\nu$) are the free parameters of our study. For the outer boundary we constructed a transmitting boundary condition, which allows matter carried by the shocks to leave the box. Fig.\ref{Parametric} summarizes our validation results and shows that the HERACLES code nicely reproduce the results of eq. (\ref{EquEta}) and (\ref{EtaAsymptotic}) . These calculations have been performed for a set of 3 initial wavelength ranging from L/4 to L/20, where L is the length of the computational domain, each one with three different initial velocity amplitudes (200, 100 and 50 $km\:s^{-1}$).

\section{Ignition conditions}

The analytical results from the previous section can predict quite accurately the shock strength and the dissipation at any position in the degenerate carbon oxygen matter of a white dwarf. But, ignition by itself depends on the thermonuclear properties of the plasma. Hereinafter we consider numerical simulation including the nuclear network described in section \ref{nuclear}, both planar and spherical geometry will be discuss.

\subsection{Slab geometry}

We showed that the perturbation wavelength plays a major role regarding the ignition process. Indeed, short wavelength perturbations generate a train of small amplitude shocks whose cumulative dissipation brings the temperature up to ignition, at some point in the profile. Conversely a large wavelength perturbation can trigger ignition through dissipation across one single shock. To show this, we adopted an initial small Mach number perturbation, M=0.01, corresponding to a velocity amplitude of $u_0 = 100$ $km\:s^{-1}$ and we varied the wavelength by one order of magnitude.

\begin{figure}
 \begin{center}
 \includegraphics[width=0.45\textwidth]{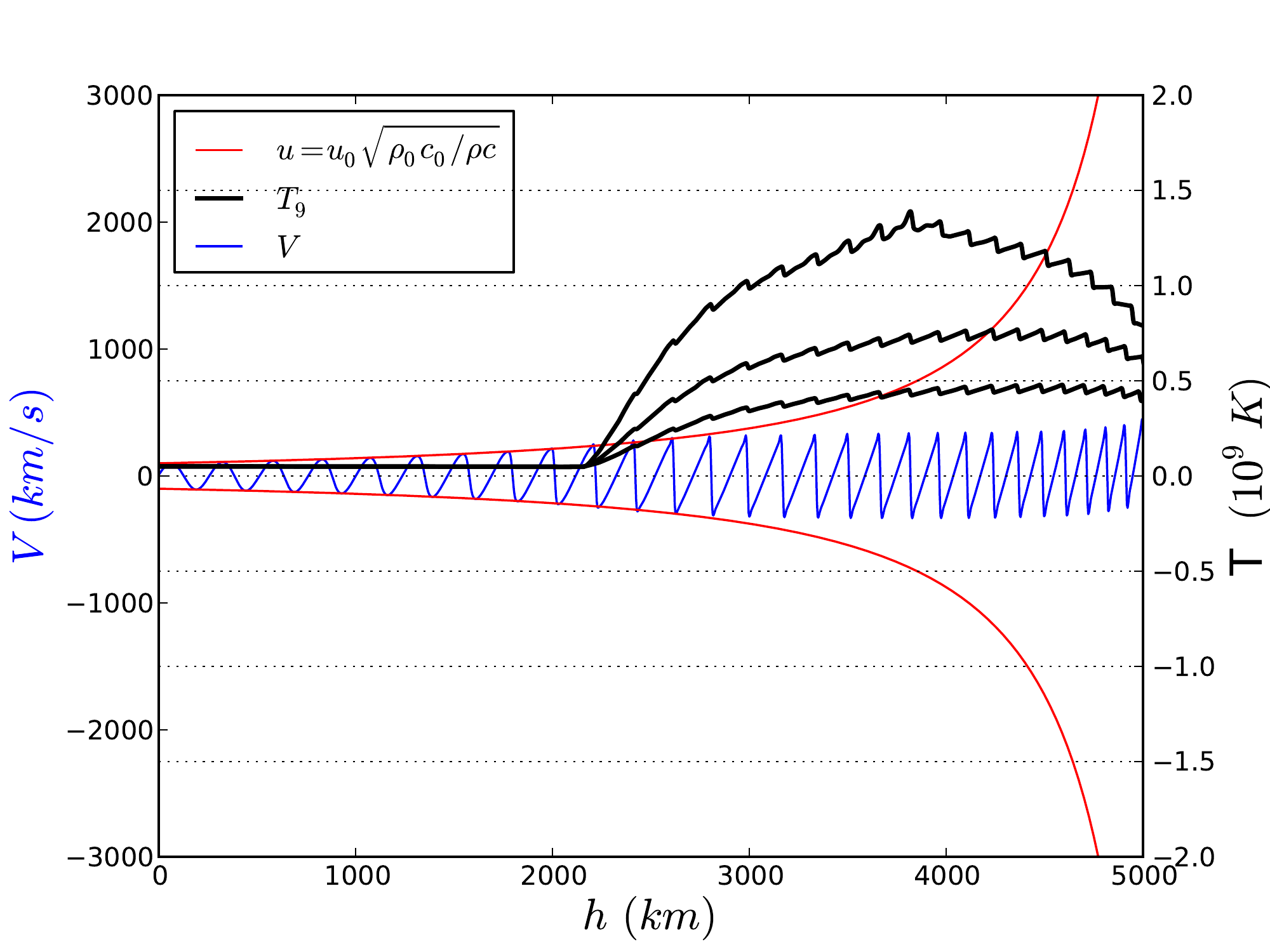}
 \caption{Propagation and amplification of a 100 $km\:s^{-1}$ or M=0.01 perturbation in the planar density gradient. In blue the velocity from the numerical simulation, compared to the amplification of eq.(\ref{Ampli_U}) in the absence of dissipation. They separate when the shocks kinetic energy is dissipated. This is stressed by the temperature at three subsequent times plotted in black.}
 \label{100-0.05}
\end{center}
\end{figure}

We consider first the short wavelength perturbation with $\lambda_0=0.05L=263$ $km$, corresponding to a frequency of 40 Hz. Transition to a shock occurs at the height of about 2500 km as predicted by eq.(\ref{ShockFormation}). This is the point where the velocity from the simulation, in blue on Fig.\ref{100-0.05}, diverges from the prediction of eq.(\ref{Ampli_U}) in red, since it enters the dissipation dominated regime. At that point the wave begins to heat up the plasma by dissipating kinetic energy. This is illustrated by the three black curves representing the temperature at three subsequent times. In that case, heating occurs through many successive weak shocks each one contributing to a small net temperature increase. This quasi-continuous dissipation process is governed by the evolution of the shock strength in eq.(\ref{Dissipation}), nicely predicted by eq.(\ref{EquEta}) (see Fig.\ref{Parametric}). For the $\alpha$-chain thermonuclear network the runaway temperature is reached after 6.9 s (250 
periods).

\begin{figure}
 \begin{center}
 \includegraphics[width=0.45\textwidth]{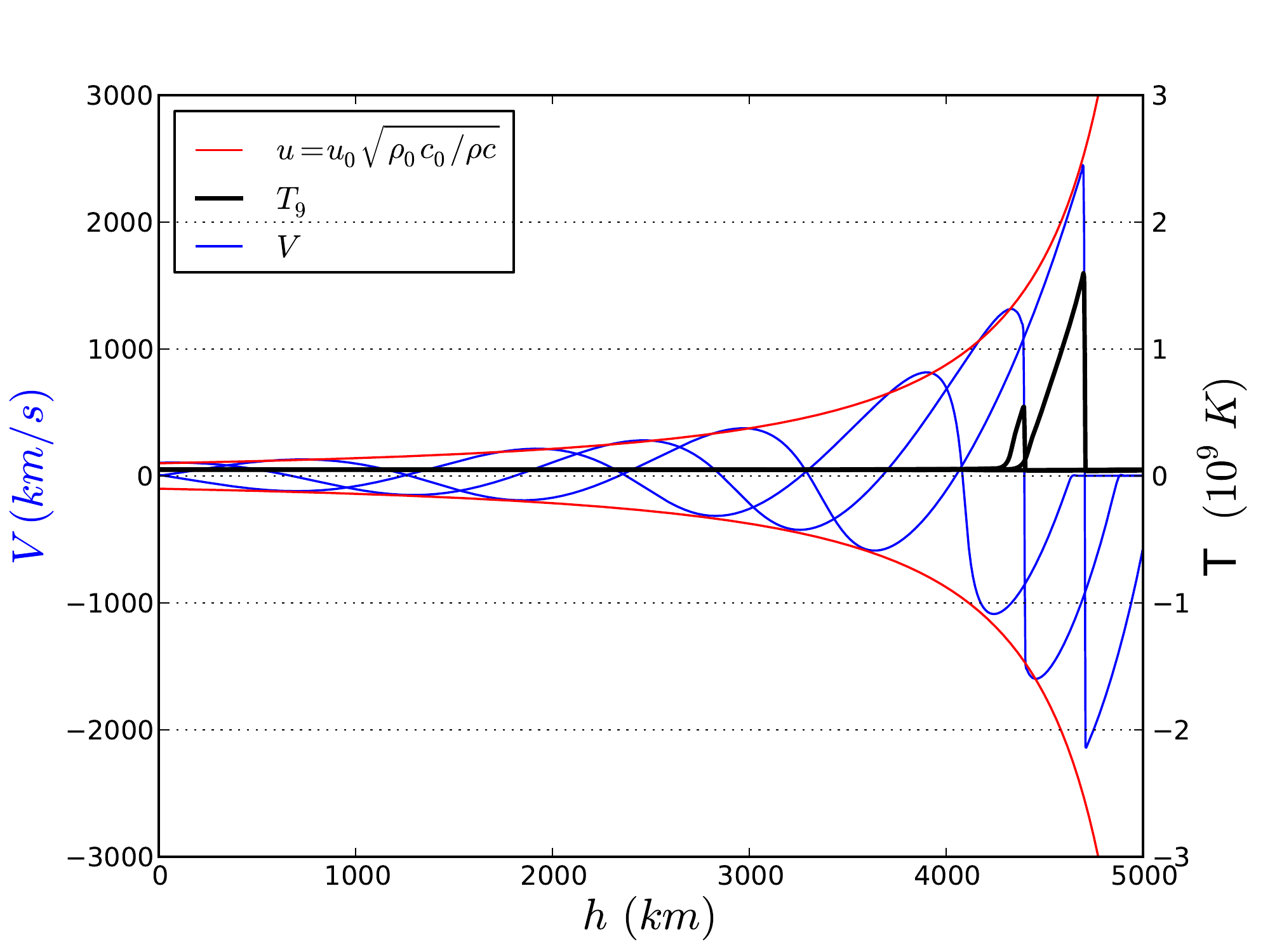}
 \caption{Propagation and amplification of the same initial perturbation in the same gradient. Here the shock is formed latter and the numerical velocity (in blue) sticks to the dissipationless prediction (in red) much longer. This results in strong dissipation as soon as the shock forms and a detonation is triggered before decompression can occur.}
 \label{100-0.5}
\end{center}
\end{figure}

We turn now to the case of larger perturbation wavelength $\lambda_0=0.5L$, with the same initial amplitude, on Fig.\ref{100-0.5}. Detonation is triggered by dissipation of a single shock, before decompression occurs. The shock forms at about $h=4500$ $km$, which again is close to the predictions of eq.(\ref{ShockFormation}). Due to a longer travel across the steep gradient, the perturbation is more amplified, according to eq.(\ref{Ampli_U}), before dissipation inhibits its growth. This leads to a much stronger shock, $ \Delta u = 4000$ $km\:s^{-1} $, and thus a very strong dissipation. On Fig.\ref{100-0.5}, when the shock forms, the post shock temperature is greater than $10^9 K$. But contrary to the short wavelength case it is immediately followed by drastic cooling in the decompression region. For ignition to occur, the post-shock induction time should be smaller than the cooling time behind the shock, which requires higher post shock temperatures. Since these single shocks are strong, this regime is not 
predictable by eq.(\ref{EquEta}), but due to long wavelength it is more easily simulated.

%

\begin{figure}
 \begin{center}
 \includegraphics[width=0.45\textwidth]{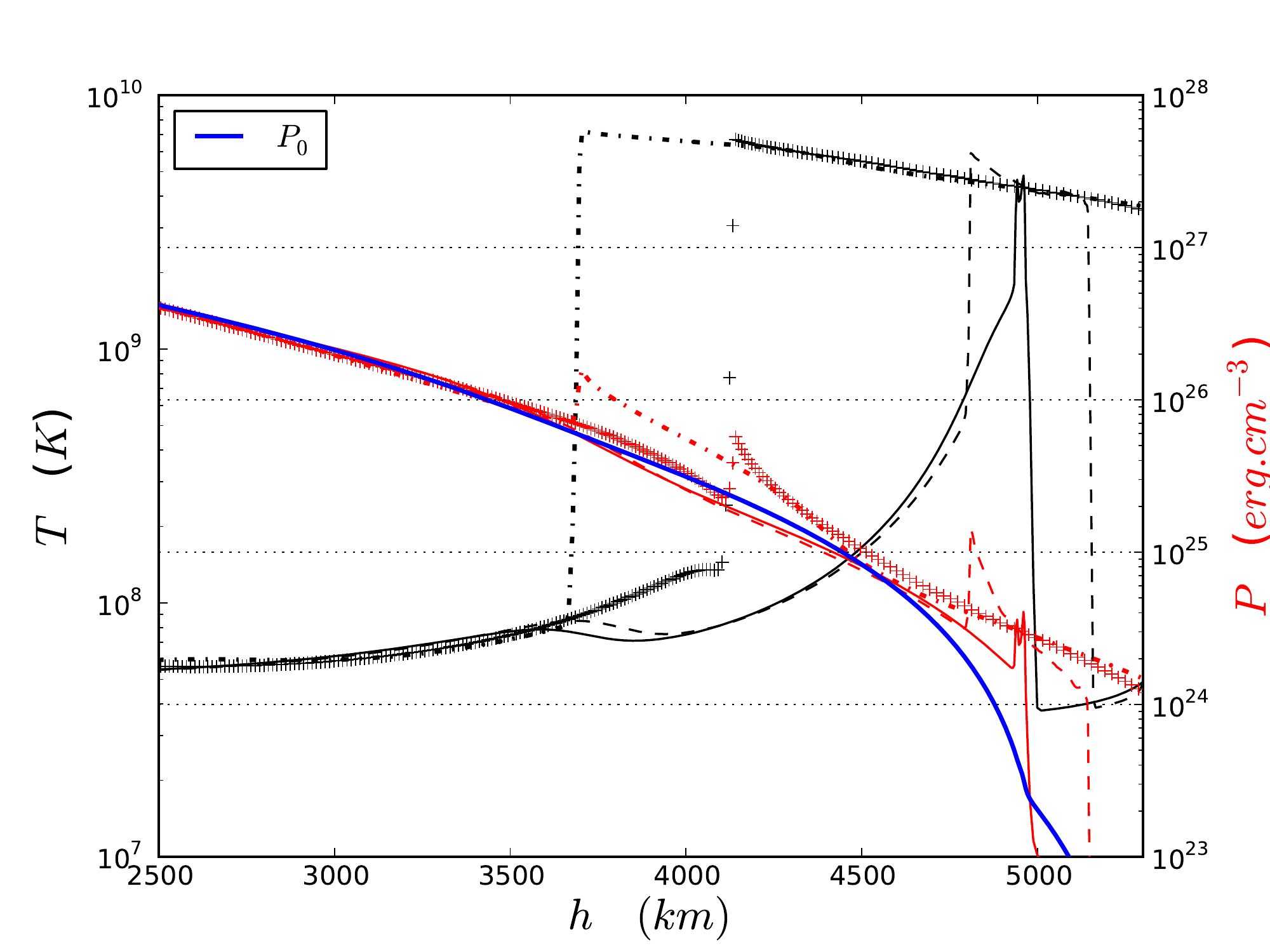}
 \caption{ Ignition and propagation of a detonation in the large wavelength case ($\lambda_0=0.5L$ and $u=100$ $km\:s^{-1}$). Temperature and pressure profiles in black and red respectively are plotted at four successive times.}
 \label{Propagation}
\end{center}
\end{figure}

Whatever the mechanism to reach ignition temperature at the hot spot, the runaway creates a strong overpressure sufficient to propagate a double detonation both upward and downward. Fig.\ref{Propagation} illustrates such a successful ignition of a detonation.

We analyse now the results of the parametric study. The number of shocks needed to ignite a detonation is presented in table \ref{Table1}, for 7 different frequencies and amplitudes varied from 200 $km\:s^{-1}$ to 30 $km\:s^{-1}$, corresponding to Mach number from 0.02 to 0.003. As discussed above increasing the initial amplitude, $u_0$, has two opposite effects. It increases the acoustic energy flux, which depends on $u_0^2$, leading to stronger shocks but dissipation occurs earlier preventing further amplification. Regarding detonation ignition, the stronger the perturbation the faster the ignition. Also at some point, when decreasing the amplitude, $H_{sh}$ enters region where the density is too low for ignition to be possible, this corresponds to the dash symbols in \ref{Table1}.

The wavelengths of the perturbations has a drastic effect as seen by comparison of the two cases described before. Although it has no effects on the energy flux ( $F=\frac{1}{2}\rho u^2 C_s$ ), it has a strong effect on the height of shock formation: higher frequency or shorter wavelength perturbations will start dissipating earlier leading to less amplification and weaker shocks. The energy flux being independent of the frequency, the same energy will diluted and deposited over wider area. On the other hand it makes it possible for weaker perturbations to trigger a detonation  given more time by reducing the shock formation height and bringing it back in denser region where ignition is possible. For example a 30 $km\:s^{-1}$ initial perturbation, would not detonate with a wavelength of $ \lambda_0 = 0.5 L$, because the shocks would form too late, but a smaller wavelength makes ignition possible after many weak shocks.

Last, the density gradient has no effect in normalised units. The parameter determining the gradient is the acceleration of gravity: g. If g is increased by a factor 10 (as is done in the lower part of table \ref{Table1}), the same density gradient will shrink to 10 times smaller box. Taking accordingly smaller wavelength the shock strengths and height formations are the same. The frequency is thus higher by a factor 10, leading to a higher rates of dissipation. But in term of dissipation per shock, it is also the same. Thus ignition of a detonation needs almost the same number of shocks. The small difference is due to smaller region at maximum temperature just after the shock. It shortens the time where reactions can run away before decompression.

\begin{table}
\begin{center}
 \begin{tabular}{|c|c|c|c|c|c|c|c|c|}
\hline
$u_0 $ ($km\:s^{-1}$) & 200 & 100 & 50  & 45  & 40  & 35 & 30 \\
\hline
$\lambda = 0.50 L$ & 1   & 1   & 2   & 2   & 3   & -  & -  \\
$\lambda = 0.33 L$ & 2   & 2   & 5   & 6   & 8   & -  & -  \\
$\lambda = 0.25 L$ & 3   & 4   & 8   & 10  & 14  & -  & -  \\
$\lambda = 0.20 L$ & 5   & 7   & 14  & 16  & 21  & 36 & -  \\
$\lambda = 0.15 L$ & 10  & 14  & 25  & 29  & 35  & 46 & -  \\
$\lambda = 0.10 L$ & 27  & 40  & 63  & 70  & 81  & 98 & 134 \\
$\lambda = 0.05 L$ & 137 & 250 & 422 & 464 & 525 &    &    \\
\hline
$10$ $\times$ $g_0$ \\
\hline
$\lambda = 0.50L$  & 1   & 1   & 3   & -   & -   & -  & - \\
$\lambda = 0.33L$  & 2   & 3   & 6   & -   & -   & -  & - \\
$\lambda = 0.25L$  & 4   & 5   & 13  & 20  & -   & -  & - \\
$\lambda = 0.20L$  & 6   & 9   & 20  & 26  & -   & -  & -  \\
$\lambda = 0.15L$  & 12  & 18  & 35  & 41  & 54 &  -  & -  \\

\hline

\end{tabular}
\caption{Parametric study of the parameter space $\lambda_{osc} \times u_0$ with the conservative code HERACLES, for the case of a shallow gradient over 5000km with $g_0= GM_\odot/(1500km)^2$ and a steeper gradient with ten times the gravity. The table shows the number of shocks needed for ignition while dash symbols represent cases where ignition is impossible because shocks form at too low density.}
\label{Table1}
\end{center}
\end{table}

\subsection{Spherical Structure}

We apply this ignition mechanism to white dwarf structures. In spherical geometry the flux is still conserved but since the surface of the spherical wave grows as $r^2$ the perturbation growth is weakened. The conservation equation now reads:

$$\rho u^2 C_s 4\pi r^2 = \rho_0 u_0^2 C_{s,0} 4\pi r_0^2, $$

\noindent which gives for the velocity perturbation at any radii:

\begin{equation}
  u(r) = u_0 \sqrt{\frac{\rho_0 C_{s,0}}{\rho(r) C_s(r)}}\frac{r_0}{r}.
 \label{Ampli_U_Sph}
\end{equation}

\begin{figure}
 \begin{center}
 \includegraphics[width=0.45\textwidth]{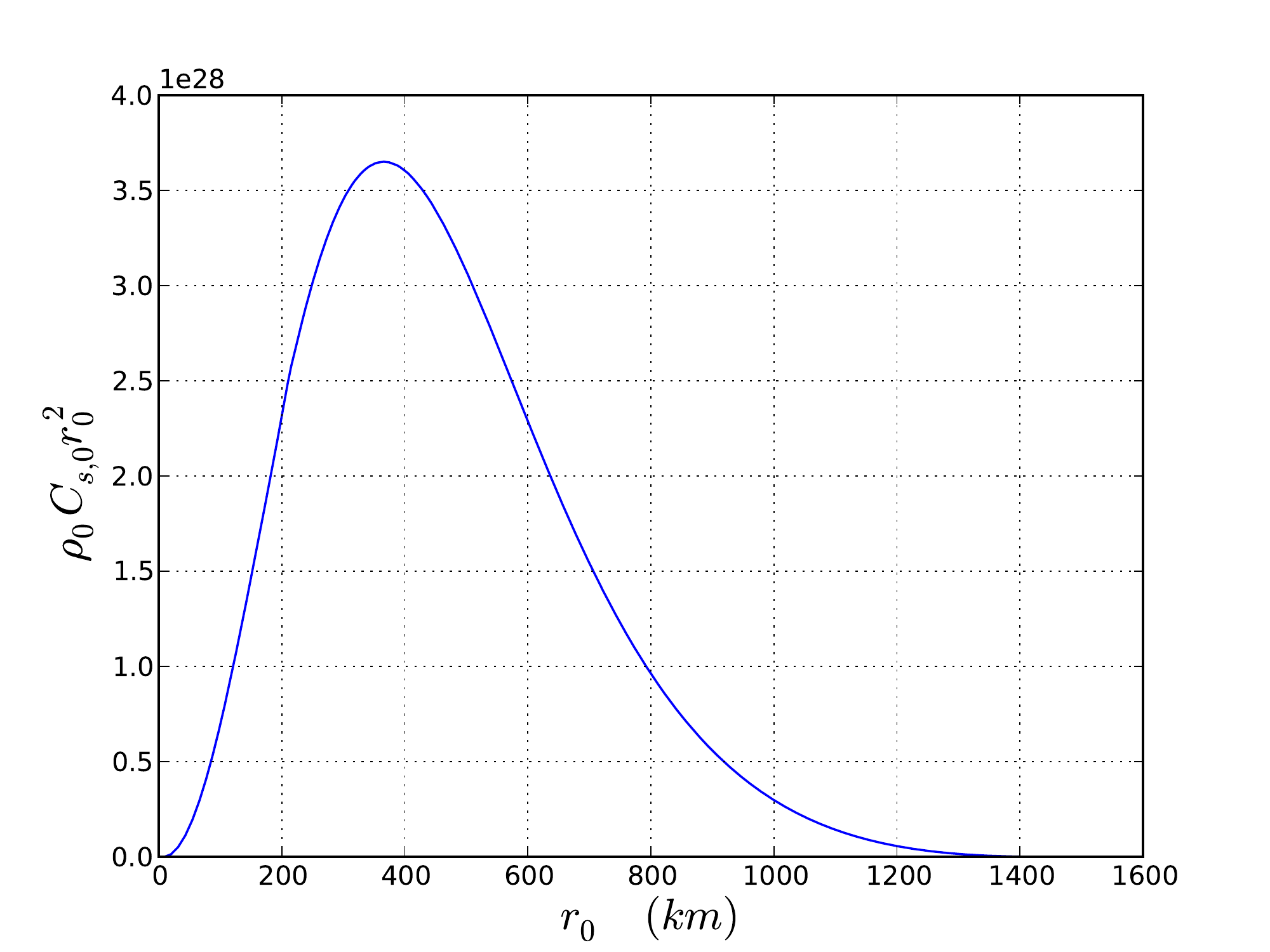}
 \caption{$\rho C_s r^2$ versus the initial perturbation position $r_0$, the maximum of this curve is the position where a perturbation will be most amplified.}
 \label{BestR0}
\end{center}
\end{figure}

\noindent Thus the amplification has to be stronger than the spherical damping $r_0/r$. This introduces a new parameter: the position $r_0 $ of the perturbation emitter. The dependence on $r_0$ of the amplification through eq.(\ref{Ampli_U_Sph}), is shown on Fig.\ref{BestR0}. At small $r_0$ geometrical damping is strong and reduce the final amplification, whereas at large radius, the variations in density and sound speed are smaller, leading to weaker amplification. The maximum of $ \rho(r_0)C_s(r_0)r_0^2 $ is the optimum position between these two opposite effects, and it lies around 350 km from the centre for the structure of the 1.43 $M_\odot$ cold C+O white dwarf used here.

The first model of supernova progenitor chosen is a cold ($T=10^7$ $K$) self-gravitating sphere of carbon and oxygen ($X_O = X_C = 0.5$). Two different models are constructed with a total mass of 1.43 $M_\odot$, with the same fixed outer pressure ($P_{ext}= 1,7.10^{21}$ $erg\:cm^{-3}$) which correspond to a density of $\rho\sim 10^5$ $g\:cm^{-3}$, but one with only carbon and oxygen and the other with a thin helium layer at the surface. The central density is $\rho_c = 7,5.10^9  $ $g\:cm^{-3}$ and sound speed: $C_s=12000$ $km\:s^{-1}$  with a radius of $ R_{WD}= 1400 $  $km$. We choose such a dense structure because it presents a very strong density gradient, best emphasizing the amplification mechanism. We will study more realistic structures of pre-expanded white dwarfs in  next section. To initiate perturbation on this spherical model, the radius $r_0$, of the oscillating pressure prescription, generating the initial perturbation is chosen. This could correspond to the position of the turbulent 
flame which would 
create pressure perturbations. At the optimal radius of 350 km the density is $2 .10^9 g\:cm^{-3}$ and the sound speed is $ C_s=10000$ $km\:s^{-1}$.

\begin{figure}
 \begin{center}
 \includegraphics[width=0.45\textwidth]{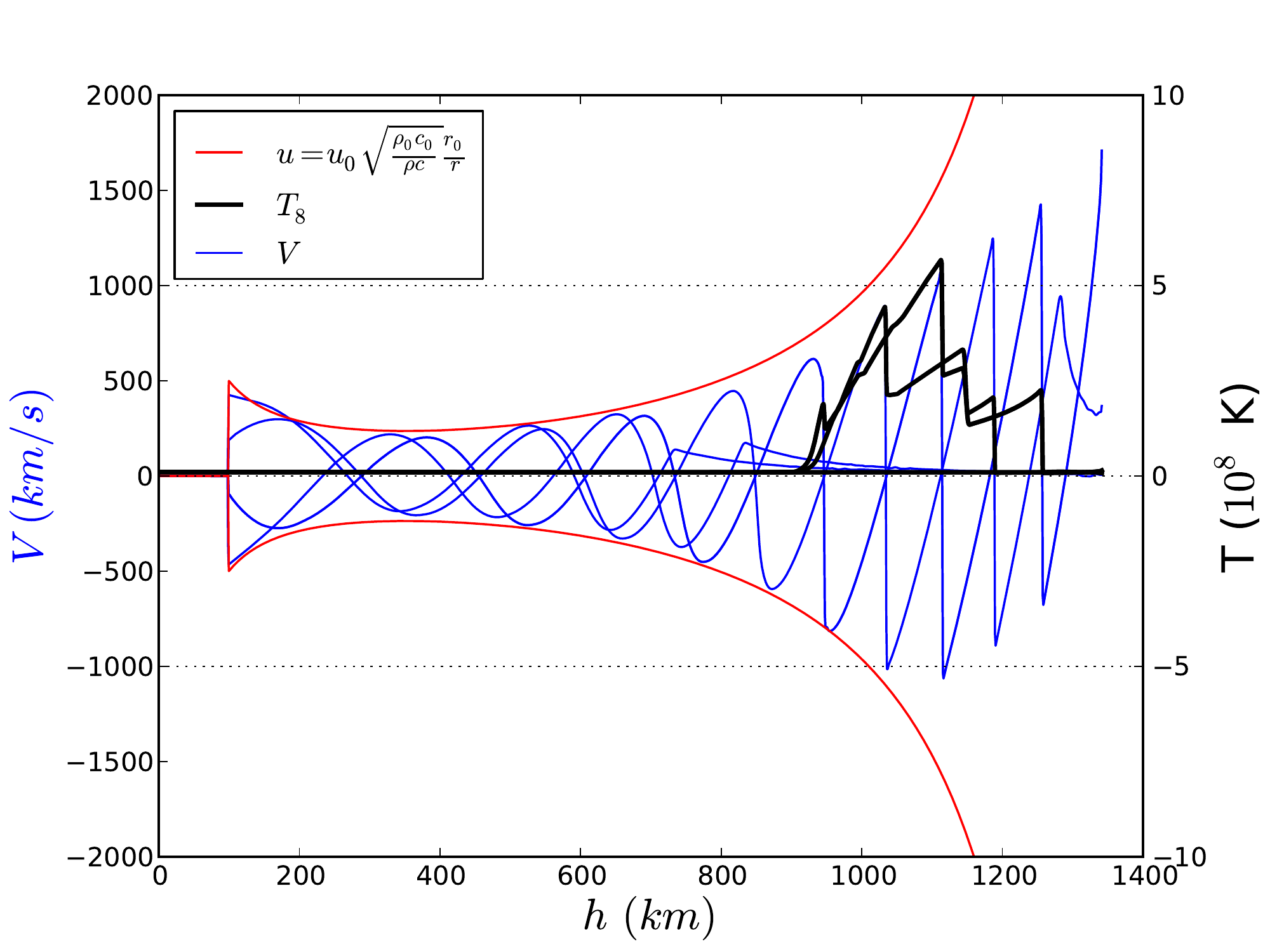}
 \caption{Propagation of a 500 $km\:s^{-1}$ perturbation of period 36 ms initiated at a $r_0=100$ $km$. The period corresponds to a wavelength of $\lambda_0=0.25 R_{WD}$ at the optimal radius: $r_0=350$ $km$. Initially geometry dampens the wave but then the gradient steepens and it is amplified. We compare the theoretical formula (\ref{Ampli_U_Sph}) for a dissipationless case to the simulation.}
 \label{WaveAmplSph}
\end{center}
\end{figure}

On Fig.\ref{WaveAmplSph}, the result for an initial perturbation of 500 $km\:s^{-1}$ ($M\sim 0.05$) originating close to the centre ($r_0=100$ $km$) is shown to emphasize the geometrical effect in the central region. Our simulation, in blue, is compared to the predictions of formula (\ref{Ampli_U_Sph}) in red. Initially where the curvature effect is strong, the damping is more important than the amplification due to the relatively flat gradient in the central region, and the amplitude of the perturbation cannot grow. But then, the gradient steepens while the effect of growing wave surface becomes smaller and the waves are amplified.

As in the planar case, we studied the possibility of detonation ignition for initial perturbations of different wavelengths and amplitudes. Here we consider the period of the waves, because the wavelength is varying with the position $r_0$. The results are shown in table \ref{Table2} and we observe the same trends. However, due to spherical damping the amplification is reduced, so that the formation of shocks strong enough to ignite a detonation necessitates larger amplitudes. In the end a quite large initial perturbation is needed: $u_0=200$ $km\:s^{-1}$, corresponding to a Mach number of $M_0 = 0.02$. We show in the following that if an helium layer is considered, the initial perturbation may be reduced to 30 $km\:s^{-1}$ or $M_0=0.003$.

\begin{table}
\begin{center}
 \begin{tabular}{|c|c|c|c|c|c|c|c|}
\hline
$u_0 $ ($km\:s^{-1}$)   & 1000 & 500 & 400 & 300  & 200 & 100  \\
\hline
$T_{osc} = 72 \; ms $ & 1 & 2   & 3   & 4    & -    & - \\
$T_{osc} = 36\; ms $  & 2 & 5   & 7   & 10   & -    & - \\
$T_{osc} = 18 \;ms $  & 4 & 13  & 17  & 34   & 67   & - \\
$T_{osc} = 9 \;ms $   & 7 & 34  & 78  & 154  & 438  &   \\

\hline

$+ M_{He} = 0.01 M_\odot$  \\

\hline
$\delta_u $ ($km\:s^{-1}$)   & 1000 & 500  & 250  & 100  & 50  & 25 \\
\hline
$T_{osc} = 72 \;ms $        &  1   &  1   &  1   &  1   & -   & -   \\
$T_{osc} = 36 \;ms $        &  2   &  1   &  1   &  3   & 5   & 20  \\
$T_{osc} = 18 \;ms $        &  4   &  5   &  6   &  9   & 19  & 78  \\
$T_{osc} = 10 \;ms $        &  7   &  34  &  35  &  47  &     &    \\

\hline
\end{tabular}

\caption{Parametric study of the parameter space $T_{osc} \times u_0$ with the conservative code HERACLES, for the case of a cold C+O white dwarf (upper table) and 0.01 $M_\odot$ helium layer (lower table). Perturbations are initiated at the optimal radius ($r_0=350$ $km$), corresponding to the maximum on Fig.\ref{BestR0}. The period $T_{osc}= 72 \;ms$ corresponds to a wavelength of $ \lambda_0 = 0.5 R_{WD} $ at the optimal radius of 350 km. Numbers in the table correspond to the number of shocks needed for detonation ignition.}
\label{Table2}
\end{center}
\end{table}

\paragraph{Helium Layer:}

The presence of an helium layer at the top of the C+O core favours a detonation. This layer is small so as to keep the explosion undisturbed.
According to \cite{2011ApJ...734...38W}, the presence of an helium shell above the C+O core would lead to unobserved peculiar light curves and spectra if the shell is larger than $\sim 0.05 M_\odot$. Thus a very thin helium layer of $M_{He}=0.01 M_\odot$ is considered. Since Helium combustion is strongly exothermic this layer detonates with weaker shocks. Once ignited, when they reach the C+O core these He-detonations are then strong enough to initiate C+O-detonations. The set-up here is different from the double detonation scenario, where an accretion induced detonation is initiated at the base of the helium layer and where the transmission of the detonation to the core may need to wait for the He-detonation induced shocks to convergence inside the core and ignite the carbon. Indeed helium either ignites above the core/envelope boundary, which leave time for the inward propagating detonation to build up large enough pressure to directly trigger the C+O-detonation or ignition occurs at the transition with 
sufficiently compressed and heated carbon so as to ignite an He-outward moving detonation and an inward C+O-detonation. And even if this fails under some conditions, the usually invoked shock convergence \citep{2010A&A...514A..53F} could still trigger the C+O-detonation, with some additional delay. The results of the simulations with an helium layer are summarised in the lower part of table \ref{Table2}. Now, the ignition of detonation necessitates a smaller perturbation of 30 $km\:s^{-1}$ or $M_0=0.003$.

\paragraph{Pre-expanded structures:}  

\begin{figure}
 \begin{center}
 \includegraphics[width=0.45\textwidth]{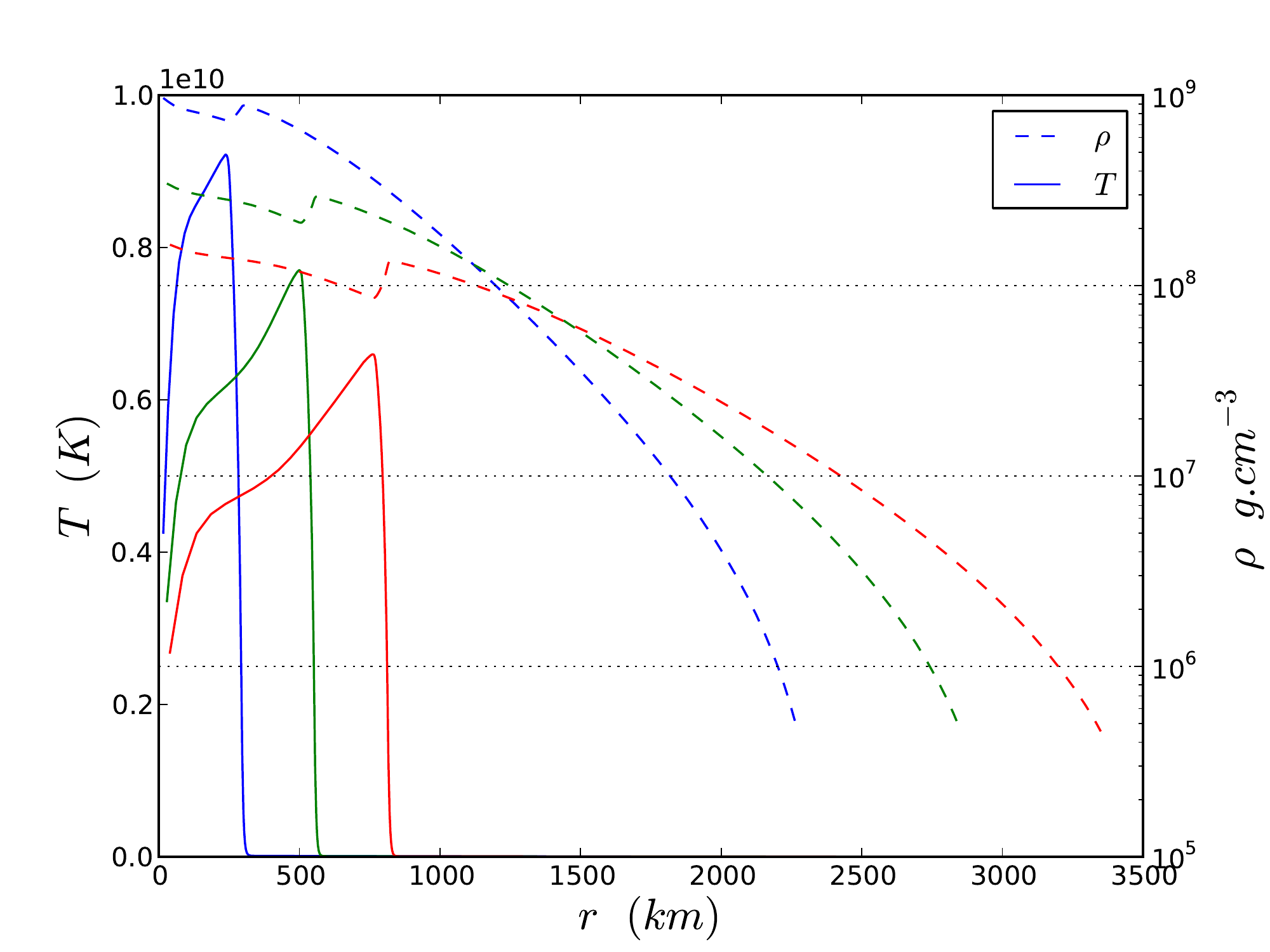}
 \caption{Initial structures of the three pre-expanded white dwarfs. The solid lines represent the temperature profiles while the dashed lines are the density profiles of these white dwarfs.}
 \label{Expanded-Structures}
\end{center}
\end{figure}

However, observations of the SN Ia nucleosynthetic yield of intermediate mass elements require that DDT should occur in a pre-expanded WD, typically 1 to 1.5 seconds after the initial thermonuclear runaway. Ideally, the previous calculations should be applied to a dynamically expanding white dwarf, with the perturbation moving with the flame front. However, the HERACLES code is presently unable to follow dynamically the expanding structure. Thus we postpone calculations with dynamic expansion for further studies. Nevertheless, as a first approach of the behaviour of the present mechanism in an expanding star, we consider three pre-expanded WD structures with densities at the flame front of $\rho_{fl}= 9\; 10^8$, $3 \; 10^8$ and $10^8$ $g\: cm^{-3}$, respectively. They are obtained with the ASTROLABE code, using its moving mesh capabilities, together with an ADR flame \citep{1995ApJ...449..695K}, to follow the whole star expansion, starting from ignition. These three structures used in the following are shown 
on Fig.\ref{Expanded-Structures}, where the temperature and density profiles are displayed. Since the densities at the perturbations initiation positions are different in these three cases, we have to consider the Mach number of the perturbations.

\begin{table}
\begin{center}
 \begin{tabular}{|c|c|c|c|c|c|c|c|}
\hline
$M$ $=$           & 0.2 & 0.15 & 0.1 & 0.05 & 0.03 & 0.02  \\

\hline
$\rho_{fl}=9.10^8$  \\

\hline
$T_{osc} = 0.15 $  & 0.3  & 0.3  & 0.3  & 0.4 & 0.5  &  0.6  \\
$T_{osc} = 0.10 $  & 0.3  & 0.3  & 0.3  & 0.5 & 0.6  &  0.7  \\
$T_{osc} = 0.05 $  & 0.3  & 0.4  & 0.5  & 0.6 & 0.9  &  1.4  \\
$T_{osc} = 0.02  $ & 0.2c & 0.4c & 1.2c &  -  &  -   &   -   \\

\hline
$\rho_{fl}=3.10^8$  \\
\hline
$T_{osc} = 0.20  $ & 0.4  & 0.4 & 0.4 & 0.8 &  1.0  & -  \\
$T_{osc} = 0.15  $ & 0.4  & 0.4 & 0.7 & 0.8 &  1.1  & -  \\
$T_{osc} = 0.10  $ & 1.0c & 0.8 & 0.8 & 0.9 &  1.1  & -  \\
$T_{osc} = 0.05  $ & 0.5c & 1.1c & -    &  -  &   -   & -  \\
$T_{osc} = 0.02  $ & 0.2c & 0.3c & 1.1c &  -  &   -   & -  \\

\hline

$\rho_{fl}=1.10^8$  \\

\hline
$T_{osc} = 0.20 $     &  0.5  &  0.6   &  1.0  & 1.0   &  -    & -   \\
$T_{osc} = 0.15 $     &  0.5  &  0.9   &  1.0  & 1.0   &  -    & -   \\
$T_{osc} = 0.10 $     &  1.3c &  1.0   &  1.2  &  -    &  -    & -   \\
$T_{osc} = 0.05  $    &  0.6c &  1.3c  &  -    &  -    &  -    & -   \\
$T_{osc} = 0.02  $    &  0.2c &  0.4c  &  1.3c &  -    &  -    & -   \\

\hline
\end{tabular}

\caption{Parametric study of the parameter space $M \times T_{osc}$, for the three cases of expanded C+O white dwarfs with an 0.01 $M_\odot$ helium layer. Perturbations are initiated at the position of the flame. $T_{osc}$ is the perturbation period in second, which varies from $0.2$ to $0.02$ $s$, while $M$ is the Mach number, in the range $0.2$ to $0.02$. Numbers in the table give the time in second necessary for detonation ignition by the perturbation of period $T_{osc}$ and Mach number $M$. When these numbers are followed by the letter c, it means ignition occurs in the carbon, otherwise ignition takes place in the helium layer.}
\label{Table3}
\end{center}
\end{table}

\noindent Due to expansion, density gradients steadily decrease, resulting in weaker and weaker  amplification of perturbations. Consequently, at later times, quite stronger perturbations are required to trigger sufficiently prompt DDT. Here we only consider detonation ignition occurring in less than 1.5 seconds, thus excluding weak perturbations. Indeed, weak perturbations must be treated in a dynamical set up, allowing for repeated shock heating from the start. Table \ref{Table3} summarizes the results for the three pre-expanded white dwarfs. For those structures the sound velocities at the flame front are $C_s=8023$, $6720$ and $5880$ $km\:s^{-1}$ respectively. To get an idea of the wavelengths of the perturbations we consider, we give the period corresponding to a wavelength of half the star radius: $T_{0.5}=0.14$, $0.21$ and $0.28$ $s$ respectively.

We find again the same trends, longer period and smaller amplitude perturbations turn to shocks farther away, so that at some point shocks form in the outer envelope, with too low density for ignition to be possible. On the opposite, shorter period and stronger perturbations start to dissipate closer to the flame, leading to ignition in the carbon-oxygen core rather than in the helium shell. These cases are marked with a letter c in the table \ref{Table3}. In this case, corresponding to high frequency perturbations, the energy is deposited mainly in the carbon-oxygen mixture and the presence of an helium layer is of no relevance. Only the strongest perturbations, reaching at least 10\% of the pressure, can trigger a timely detonation. Moreover, due to the short distance between the flame, where perturbations are generated, and the locus of shock formation, where the energy is dissipated, the ignition doesn't really rely on amplification by the density gradient, since density does not significantly change 
over such a short distance. In fact, it relies mainly upon the natural 
evolution of any perturbation to steepen into shocks, so that the ignition times are almost independent of the structure. However, since the  energy is deposited close to the flame, the fact that the flame is actually moving, may prevent the temperature to reach ignition conditions. But in any case, the preheating of fuel due to shocks will have strong effects on the flame when it reaches this pre-heated region. Conversely long period perturbations with smaller amplitudes could trigger a detonation in the helium shell, due to a longer delay for shock formation. As the star expands, the minimum amplitude to detonate the helium layer increases, due to less and less amplification through a flatter and flatter gradient. From $M=0.003$ in the cold and dense initial white dwarf, the required triggering pressure perturbation, $\delta P/P$, are respectively $0.02$, $0.03$ and finally $0.05$.

\section{Discussion}

Finally we can estimate the energy flux needed for a cold white dwarf with a thin helium layer to detonate through the amplification mechanism as:
$$ F_m = 0.5 \rho_0  C_{s,0} u^2 \sim 2.5 \; 10^{31}  \; erg \: cm^{-2} \:s^{-1}, $$

\noindent with $\rho_0=2\:10^9 \; g\:cm^{-3}$, $C_{s,0}=10000 \; km \:s^{-1}$ and a perturbation amplitude of M=0.005 ($u=50\;km\;s^{-1}$). The question is then to find how and where this flux could be produced. An estimate of the energy flux generated by the thermonuclear combustion during the initial deflagration phase at the density $\rho_0$, is:
$$ F_{nuc} = V_{fl} \rho \epsilon_{nuc} = 8.7 \; 10^{33} \;erg\:cm^{-2} \: s^{-1}.  $$

\begin{table}
\begin{center}
 \begin{tabular}{|c|c|c|c|}
\hline
$\rho_{fl}$ $(g\:cm^{-3})$  & $9\:10^8$ & $3\:10^8$ & $10^8$   \\
\hline
$F_m $ $(erg\:cm^{-2}\:s^{-1})$   & $9.3\:10^{31}$   & $4.1\:10^{31}$   & $2.5\:10^{31}$  \\
\hline
$F_{nuc}  $ $(erg\:cm^{-2}\:s^{-1})$  & $1.6\:10^{33}$   & $2.2\:10^{32}$   & $2.3\:10^{31}$  \\
\hline

\end{tabular}

\caption{Comparison of the acoustic flux needed for detonation ignition to the nuclear energy generation of the corresponding laminar flame in the three pre-expanded white dwarfs studied previously.}
\label{Flux}
\end{center}
\end{table}

\noindent The laminar flame velocity used here is: $V_{fl} = 75 \;km\:s^{-1}$ in agreement with \cite{1992ApJ...396..649T} and the nuclear energy released is: $\epsilon_{nuc}=5.8 \; 10^{17}$ $erg \: g^{-1}$, corresponding to the combustion of the $X_O=X_C=0.5$ mixture. The laminar deflagration speed is used as a lower limit since it should be greatly enhanced by turbulence. But even considering this limit, only 0.3\% of the generated energy injected into sound waves would be sufficient for a successful ignition in the cold initial white dwarf. In Table \ref{Flux} the same comparison is made for the pre-expanded white dwarfs. For these structures the required energy represents a more and more substantial part of the available nuclear energy: $6\%$, $19\%$ and finally almost all the energy generated by a laminar flame. Accordingly, only fast turbulent flames could generate, at late times, enough energy for this mechanism.

As stressed in the previous section, a strong constraint on perturbations frequencies and amplitudes would be the time scale over which the transition to detonation has to occur. In the context of the delayed detonation model of SNe Ia, the detonation has to be triggered after an initial phase of expansion driven by the energy released by the deflagration. Thus, it should not be too fast at the early stage of flame propagation, so that an early detonation is avoided. But it should be fast enough, so that the detonation can be triggered before the expansion gets too large. In the classical local DDT, many studies agree to say the detonation has to occur when the flame reaches densities of about $2\:10^7$ $g\:cm^{-3}$. In this paper the DDT mechanism is \textit{non local}, with a shift in space and time between the deflagration, producing the perturbations, and the position where they trigger a detonation. Thus the critical density, derived from usual DDT, is not directly relevant to this mechanism. Taking into account the delay due to wave propagation and shock heating accumulation, we have considered white dwarfs where the flame is still at densities above the classical critical density, so that during the shock heating period, the star will further expand and detonation will be initiated when the flame  encompasses regions with approximately the critical density. Obviously a serious drawback of this study is to consider a static flame position, in a structure devoid of previous shock formation. Indeed, shock heating should occur right from the beginning, with evolving characteristics as the flame travel across the white dwarf. The final heating will thus be a complex interplay between the moving flame and the history of shocks formation and dissipation. Thus, the next step after this preliminary study, will be to take properly into account the expansion of the star, along with a moving perturbation generator.

Finally, we found that there are two kinds of perturbations that could successfully trigger a non-local DDT. Either long period and small amplitude perturbations, corresponding to large combustion scale, like Rayleigh-Taylor mushrooms. Those perturbations will heat up the outer helium layer, igniting an helium detonation, given the right shock heating history. Or, conversely, high frequency and large amplitude perturbations, corresponding to repeated strong pressure perturbations, which could be associated with small scale turbulent combustion. In that case it is unlikely that there will be enough time for shock heating to reach ignition temperature, since shocks will form not so far from the flame. But in case of detonation ignition failure, the fuel would have been efficiently pre-heated, which would have a positive feedback on the flame.

In the classical DDT through the gradient mechanism of \cite{1991A&A...246..383K}, pressure perturbation has to build up directly to detonation pressure, and this necessitate over 10 km. In our case a failed build up to only a fraction of that pressure would be enough, if sufficiently repeated. Moreover, in an attempt to get direct transition to detonation, Poludnenko (private communication) obtained strong pressure pulses of about 30\% of the unperturbed pressure, repeated every eddy turn-over time. This perturbation corresponds to column 2 of table \ref{Table3} and might detonate, given the right conditions. To conclude, we propose a promising possibility for the revival of failed local DDT, as non-local DDT induced by density gradient amplification.

\bibliographystyle{aa}
\bibliography{Article}

\begin{thebibliography}{31}
\expandafter\ifx\csname natexlab\endcsname\relax\def\natexlab#1{#1}\fi

\bibitem[{{Arnett}(1969)}]{1969Ap&SS...5..180A}
{Arnett}, W.~D. 1969, \apss, 5, 180

\bibitem[{{Brinkley} \& {Kirkwood}(1947)}]{1947PhRv...72.1109B}
{Brinkley}, S.~R. \& {Kirkwood}, J.~G. 1947, Physical Review, 72, 1109

\bibitem[{Ciccarelli \& Dorofeev(2008)}]{Ciccarelli2008499}
Ciccarelli, G. \& Dorofeev, S. 2008, Progress in Energy and Combustion Science,
  34, 499

\bibitem[{{Colella} \& {Glaz}(1985)}]{1985JCoPh..59..264C}
{Colella}, P. \& {Glaz}, H.~M. 1985, Journal of Computational Physics, 59, 264

\bibitem[{{Fink} {et~al.}(2010){Fink}, {R{\"o}pke}, {Hillebrandt},
  {Seitenzahl}, {Sim}, \& {Kromer}}]{2010A&A...514A..53F}
{Fink}, M., {R{\"o}pke}, F.~K., {Hillebrandt}, W., {et~al.} 2010, \aap, 514,
  A53

\bibitem[{{Fryxell} {et~al.}(2000){Fryxell}, {Olson}, {Ricker}, {Timmes},
  {Zingale}, {Lamb}, {MacNeice}, {Rosner}, {Truran}, \&
  {Tufo}}]{2000ApJS..131..273F}
{Fryxell}, B., {Olson}, K., {Ricker}, P., {et~al.} 2000, ApJS, 131, 273

\bibitem[{{Gamezo} {et~al.}(2005){Gamezo}, {Khokhlov}, \&
  {Oran}}]{2005ApJ...623..337G}
{Gamezo}, V.~N., {Khokhlov}, A.~M., \& {Oran}, E.~S. 2005, \apj, 623, 337

\bibitem[{{Gonz{\'a}lez} {et~al.}(2007){Gonz{\'a}lez}, {Audit}, \&
  {Huynh}}]{2007A&A...464..429G}
{Gonz{\'a}lez}, M., {Audit}, E., \& {Huynh}, P. 2007, \aap, 464, 429

\bibitem[{{Iben} \& {Tutukov}(1984)}]{1984ApJS...54..335I}
{Iben}, Jr., I. \& {Tutukov}, A.~V. 1984, \apjs, 54, 335

\bibitem[{{Ivanov} {et~al.}(2010){Ivanov}, {Kiverin}, {Liberman}, \&
  {Fortov}}]{2010DokPh..55..480I}
{Ivanov}, M.~F., {Kiverin}, A.~D., {Liberman}, M.~A., \& {Fortov}, V.~E. 2010,
  Physics - Doklady, 55, 480

\bibitem[{{Jackson} {et~al.}(2010){Jackson}, {Calder}, {Townsley}, {Chamulak},
  {Brown}, \& {Timmes}}]{2010ApJ...720...99J}
{Jackson}, A.~P., {Calder}, A.~C., {Townsley}, D.~M., {et~al.} 2010, \apj, 720,
  99

\bibitem[{Khokhlov {et~al.}(1999)Khokhlov, Oran, \& Thomas}]{Khokhlov1999323}
Khokhlov, A., Oran, E., \& Thomas, G. 1999, Combustion and Flame, 117, 323

\bibitem[{{Khokhlov}(1991{\natexlab{a}})}]{1991A&A...245..114K}
{Khokhlov}, A.~M. 1991{\natexlab{a}}, \aap, 245, 114

\bibitem[{{Khokhlov}(1991{\natexlab{b}})}]{1991A&A...246..383K}
{Khokhlov}, A.~M. 1991{\natexlab{b}}, \aap, 246, 383

\bibitem[{{Khokhlov}(1995)}]{1995ApJ...449..695K}
{Khokhlov}, A.~M. 1995, \apj, 449, 695

\bibitem[{{Kushnir} {et~al.}(2012){Kushnir}, {Livne}, \&
  {Waxman}}]{2012ApJ...752...89K}
{Kushnir}, D., {Livne}, E., \& {Waxman}, E. 2012, \apj, 752, 89

\bibitem[{{Landau} \& {Lifshitz}(1959)}]{1959flme.book.....L}
{Landau}, L.~D. \& {Lifshitz}, E.~M. 1959, {Fluid mechanics}

\bibitem[{{Lighthill}(1955)}]{1955IAUS....2..121L}
{Lighthill}, M.~J. 1955, in IAU Symposium, Vol.~2, Gas Dynamics of Cosmic
  Clouds, 121

\bibitem[{Oran \& Gamezo(2007)}]{Oran20074}
Oran, E.~S. \& Gamezo, V.~N. 2007, Combustion and Flame, 148, 4

\bibitem[{{Osterbrock}(1961)}]{1961ApJ...134..347O}
{Osterbrock}, D.~E. 1961, \apj, 134, 347

\bibitem[{{Pakmor} {et~al.}(2012){Pakmor}, {Kromer}, {Taubenberger}, {Sim},
  {R{\"o}pke}, \& {Hillebrandt}}]{2012ApJ...747L..10P}
{Pakmor}, R., {Kromer}, M., {Taubenberger}, S., {et~al.} 2012, \apjl, 747, L10

\bibitem[{{Poludnenko} {et~al.}(2011){Poludnenko}, {Gardiner}, \&
  {Oran}}]{2011PhRvL.107e4501P}
{Poludnenko}, A.~Y., {Gardiner}, T.~A., \& {Oran}, E.~S. 2011, Physical Review
  Letters, 107, 054501

\bibitem[{{R{\"o}pke} {et~al.}(2007){R{\"o}pke}, {Hillebrandt}, {Schmidt},
  {Niemeyer}, {Blinnikov}, \& {Mazzali}}]{2007ApJ...668.1132R}
{R{\"o}pke}, F.~K., {Hillebrandt}, W., {Schmidt}, W., {et~al.} 2007, \apj, 668,
  1132

\bibitem[{{Timmes}(1999)}]{1999ApJS..124..241T}
{Timmes}, F.~X. 1999, \apjs, 124, 241

\bibitem[{{Timmes} \& {Swesty}(2000)}]{2000ApJS..126..501T}
{Timmes}, F.~X. \& {Swesty}, F.~D. 2000, ApJS, 126, 501

\bibitem[{{Timmes} \& {Woosley}(1992)}]{1992ApJ...396..649T}
{Timmes}, F.~X. \& {Woosley}, S.~E. 1992, \apj, 396, 649

\bibitem[{{Ulmschneider}(1967)}]{1967ZA.....67..193U}
{Ulmschneider}, P. 1967, \zap, 67, 193

\bibitem[{{Ulmschneider}(1970)}]{1970SoPh...12..403U}
{Ulmschneider}, P. 1970, \solphys, 12, 403

\bibitem[{{Whelan} \& {Iben}(1973)}]{1973ApJ...186.1007W}
{Whelan}, J. \& {Iben}, Jr., I. 1973, \apj, 186, 1007

\bibitem[{{Woosley} \& {Kasen}(2011)}]{2011ApJ...734...38W}
{Woosley}, S.~E. \& {Kasen}, D. 2011, \apj, 734, 38

\bibitem[{{Woosley} {et~al.}(2009){Woosley}, {Kerstein}, {Sankaran}, {Aspden},
  \& {R{\"o}pke}}]{2009ApJ...704..255W}
{Woosley}, S.~E., {Kerstein}, A.~R., {Sankaran}, V., {Aspden}, A.~J., \&
  {R{\"o}pke}, F.~K. 2009, \apj, 704, 255

\end{thebibliography}

\end{document}